\documentclass[aps,prl,reprint,showpacs,superscriptaddress,longbibliography,nourl,noeprint]{revtex4-2}
\usepackage{amsmath,amssymb,physics,bm}
\usepackage{graphicx, microtype} 
\usepackage[colorlinks=true,linkcolor=blue,citecolor=blue,urlcolor=blue]{hyperref}
\usepackage[all]{hypcap} 
\usepackage[colorinlistoftodos]{todonotes}
\usepackage{graphicx} 

\newcommand{\rvec}[0]{{\bm r}} 
\newcommand{\Evec}[0]{{\bm E}} 
\newcommand{\nvec}[0]{{\bm n}}

\begin{document}


\title[]{Coarse grained model for skyrmion dynamics}

\author{Tomás Alvim}
\affiliation{Centro de Física Teórica e Computacional, Faculdade de Ciências, Universidade de Lisboa, 1749-016 Lisboa, Portugal.}
 \affiliation{Departamento de Física, Faculdade de Ciências, Universidade de Lisboa, P-1749-016 Lisboa, Portugal.}
\author{Margarida M. Telo da Gama}%
  \affiliation{Centro de Física Teórica e Computacional, Faculdade de Ciências, Universidade de Lisboa, 1749-016 Lisboa, Portugal.}
 \affiliation{Departamento de Física, Faculdade de Ciências,
Universidade de Lisboa, P-1749-016 Lisboa, Portugal.}
\author{Mykola Tasinkevych}
\email{mykola.tasinkevych@ntu.ac.uk}
  \affiliation{Centro de Física Teórica e Computacional, Faculdade de Ciências, Universidade de Lisboa, 1749-016 Lisboa, Portugal.}
 \affiliation{Departamento de Física, Faculdade de Ciências,
Universidade de Lisboa, P-1749-016 Lisboa, Portugal.}
\affiliation{SOFT Group, School of Science and Technology, Nottingham Trent University, Clifton Lane, Nottingham NG11~8NS, United Kingdom.}
\affiliation{International Institute for Sustainability with Knotted Chiral Meta Matter, Hiroshima University, Higashihiroshima  739-8511, Japan.}

\date{\today}

\begin{abstract}
\noindent
Liquid crystal skyrmions are topologically protected spatially localized distortions of the director field which are fascinating from both fundamental mathematics and applied physics points of view. Skyrmions are realized in experimental setups which are identical to those used in display technology. This opens exciting opportunities for designing advanced electro-optical applications. Skyrmions exhibits particle-like properties including translational motion in time periodic electric fields. Despite a large volume of experimental results on the skyrmion collective behaviour, the theoretical understanding of their effective interactions and emerging dynamics is rather limited due to computational heaviness of numerical models based upon fine grained Frank-Oseen or Landau-de Gennes continuum approaches. Here, we develop a coarse grained model of the skyrmion dynamics. The model reveals the underpinning physical mechanism of the skyrmion motion driven by pulse width modulated electric fields. The mechanism is based upon the complex dynamics of the width of the twist wall around the skyrmion's core. The width evolves in a non-reciprocal way within each ${\mathbf{on}}$ and ${\mathbf{off}}$ states of the field, resulting in a net displacement of the skyrmion. We analyse in details the average skyrmion velocity as a function of the frequency and strength of the field as well as its duty cycle. The model predictions agrees qualitatively with the results of experiments and full numerical minimization of the fine grained models. Our results provide insights into the conditions necessary to observe velocity reversal as a function of the field parameters.  
\end{abstract}

\maketitle

\section{Introduction}

 Recent experimental \cite{Smalyukh2010,Ackerman2014,Ackerman2017,Sohn2018,Sohn2019a,Sohn2019,Song2021} and theoretical \cite{Bogdanov2003,Duzgun2018,Duzgun2021,Duzgun2022,Coelho2022,Long2021} studies have demonstrated that chiral nematic liquid crystals (LC) can host a large variety of topologically protected localized solitonic configurations. Analogously to standard solitons they are characterised by certain topological invariants which do not change by continuous morphing of the underlying director field. These quasi-particle excitations are dubbed skyrmions, as low-dimensional analogs of Skyrme solitons in nuclear physics \cite{Skyrme1962}. The topology and structure of many director configurations found in chiral LCs  mimic those in chiral magnets \cite{Muhlbauer:2009,Yu2010,Zhang2016,Liu:2018,Sutcliffe_2018,Das2019}, which in these materials have relevance in applications such as spintronic memory storage devices \cite{Krause2016}.

In confined geometries, LC skyrmions can be brought into controlled directional motion by time dependent external electric fields \cite{Ackerman2017,Sohn2018}. The skyrmion velocity sensitively depends on the frequency, strength  and the duty cycle of the pulse width modulated electric field. In multi-skyrmion systems, remarkable collective behaviour has been reported experimentally including light controlled skyrmion interactions and self-assembly \cite{Sohn2019a}, reconfigurable cluster formation and formation of large-scale skyrmion crystals mediated by out-of-equilibrium elastic interactions \cite{Sohn2019}.

Driven LC skyrmions offer a new paradigm of solitonic active particle-like structures without mass transport \cite{Ackerman2017,Sohn2018}. The experimental setups employed to stabilise and study active skyrmions are similar to those used in LC display technologies \cite{Shen2020}. As such, the controlled motion of a large number of skyrmions holds potential for the development of novel electro-optic responsive materials \cite{Wu2022}. Numerical analysis based on minimization of the Frank-Oseen \cite{Ackerman2017,Sohn2018} and Landau-de Gennes \cite{Duzgun2018,Duzgun2021,Duzgun2022}  free energies have successfully reproduced many experimental results regarding the structure and dynamics of skyrmionic excitations. These fine grained approaches fully resolve the spatio-temporal structure of the LC order parameter field and are computationally costly, which does not allow a comprehensive sampling of the whole space of the model parameters. 

To overcome this challenge, Long and Selinger proposed a coarse grained model of the skyrmion dynamics \cite{Long2021}, where a few macroscopic degrees of freedom effectively specify the skyrmion 
configuration and its response to an external field. The method was applied to characterise the motion of one dimensional sine-Gordon solitons, and its extension to two dimensional (2D) skyrmions was outlined without detailed analysis. 

Here, based on the original idea \cite{Long2021} we have developed a coarse grained model of driven motion of LC skyrmions in 2D. The developed model uncovers the underpinning physical mechanism of the net displacement of the skyrmion under periodic switching of the electric field ${\mathbf{on}}$ and ${\mathbf{off}}$. The mechanism is related to the complex dynamics of the width of the twist wall around the skyrmion. The width changes in a non-reciprocal way within each ${\mathbf{on}}$ and ${\mathbf{off}}$ state of the field, resulting in the net skyrmion displacement over one period of the electric field. Skyrmion velocity is proportional to the time derivative of the polar angle specifying the far field director, which responds to the changes in the applied voltage, and the speed is maximal just after the electric field has been turned ${\mathbf{on}}$ or ${\mathbf{off}}$. The analysis also demonstrates that the possibility for velocity reversal by changing the field frequency or the duty cycle is related to the ratio of the director relaxation times during the field-$\mathbf{on}$ and field-$\mathbf{off}$ states. The relaxation time is given by the inverse square of the external effective field, which also includes a fictitious component mimicking the effects of strong boundary conditions in real 3D systems.     

 \section{Model}
\subsection{Skyrmion Ansatz} 
\begin{figure}[h]
    \centering
    \includegraphics[width = 0.9\linewidth]{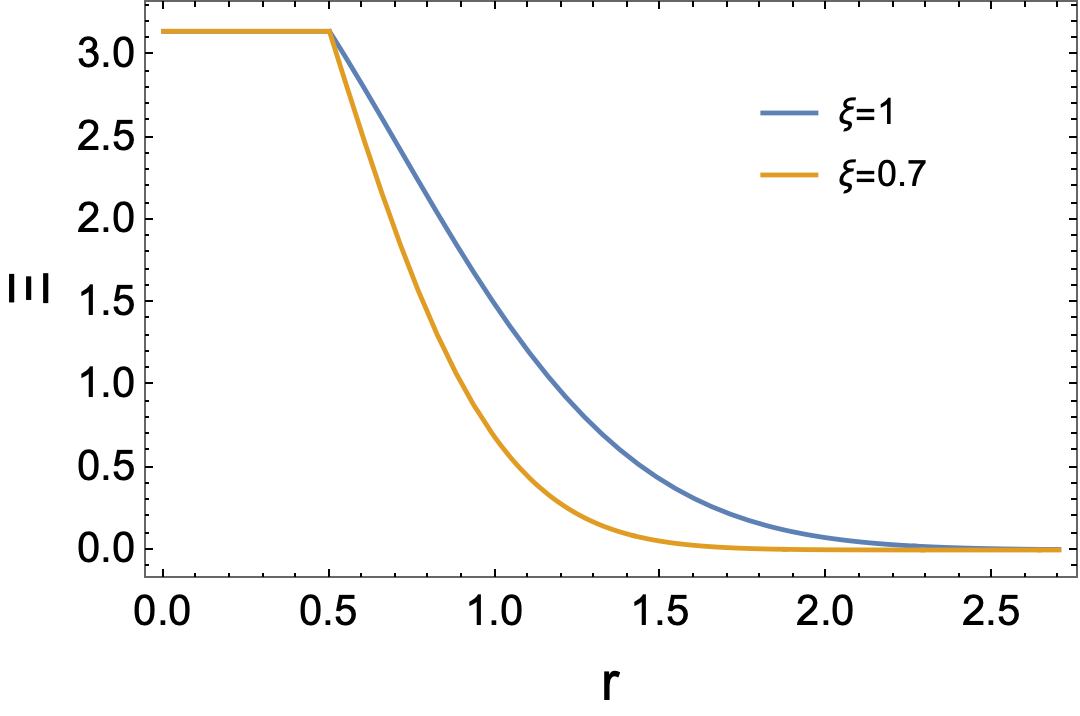}
    \caption{{\bf Polar angle of the director Ansatz for a skyrmion}. $\Xi(r;\xi)$ defined in Eq.~(\ref{eq:directo_polar_angle}) as a function of $r$, for two values of the wall thickness $\xi=p$ and $0.7p$.}
    \label{fig:polar director}
\end{figure}

 We assume that at zero electric field the far field nematic director is aligned along the $z-$axis, and define an axisymmetric {\it Ansatz} for the  skyrmion configuration with the winding number equal unity as follows:

\begin{eqnarray}
 && n^a_x(\rvec;\xi) = \sin \left ( \Xi(r;\xi) \right ) \cos \left ( \Psi(r) \right ) \nonumber\\
 && n^a_y(\rvec;\xi) = \sin \left ( \Xi(r;\xi) \right ) \sin \left ( \Psi(r) \right ) \nonumber\\
 && n^a_z(\rvec;\xi) = \cos \left ( \Xi(r;\xi) \right ),
 \label{eq:ansatz n0}
\end{eqnarray}
where
\begin{eqnarray}
 &&\Xi(r;\xi)=\pi + \pi H(\frac{p}{2}-r)
 \left[ \exp \left(\frac{(r-\frac{p}{2})^2}{\xi^2}\right) -1\right] \\
 \label{eq:directo_polar_angle}
 && \Psi(r) = \tan^{-1}\left(\frac{x-x_s}{y-y_s}\right) + \frac{\pi}{2}\\
 && r = \sqrt{(x-x_s)^2+(y-y_s)^2}.
\end{eqnarray}
Here $H$ is the Heaviside step function, $p$ is the cholesteric pitch and $r$ is the distance from a point $(x,y)^{T}$ to the skyrmion center at $(x_s,y_s)^T$, where the superscript $T$ denotes the transposition operation. The parameter $\xi$ controls the width of the twist wall which separates the skyrmion core of size $p/2$ with the director pointing in $-\hat{z}$ direction from the far field director $\hat{z}$. The function $\Xi(r;\xi)$ is plotted in Fig.~\ref{fig:polar director}.

\begin{figure}[h]
\centering
    \includegraphics[width = \linewidth]{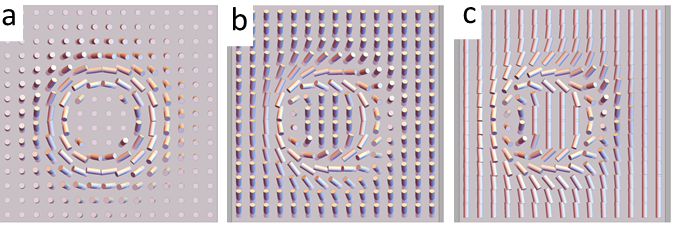}
        \caption{{\bf Director profile of the skyrmion Ansatz}. The director field of the skyrmion configurations for $\Theta = 0, \frac{\pi}{4},\frac{\pi}{2}$ (a), (b), and (c), respectively; $\Phi = 0$ in all the cases. The director field is represented by cylinders. The skyrmion core has the same alignment as the background. The wall thickness parameter is $\xi=0.7p$.}
         \label{fig:rod_snapshots}
\end{figure}

In this study we consider LCs with positive dielectric anisotropy $\Delta \varepsilon$, and thus when an external electric field is applied perpendicular to $\hat{z}$, the skyrmion will morph to a non-symmetic so-called bimeron \cite{Li2020} configuration, with the far field director $\nvec_{0}=(\sin\Theta\cos\Phi,\sin\Theta\sin\Phi,\cos\Theta)^T$ tilted away from $\hat{z}$. For strong enough fields $\nvec_{0}$ will be in the $(x,y)$ plane. The resulting bimeron configuration has no analytical representation, therefore we resorted to a simple approximation, which consists in a uniform local rotation of the symmetric director field in Eq.~(\ref{eq:ansatz n0}) 
\begin{equation}
\nvec(\rvec; \Theta,\Phi, \xi )= R(\Theta,\Phi) \, \nvec^a( \rvec;\xi ).
\label{eq:R2}
\end{equation}
Here $R(\Theta,\Phi)$ is the rotation matrix which transforms $\hat{z}$ to $\nvec_{0}$. Figure~\ref{fig:rod_snapshots}(a) illustrates the director configuration of the original {\it Ansatz} in Eq.~(\ref{eq:ansatz n0}), and Figs.~\ref{fig:rod_snapshots}(b) and (c) show the approximation (\ref{eq:R2}) of the bimeron configurations for $\Theta = \pi/4$ and $\pi/2$.

One way to obtain $\nvec_{0}$ is to rotate $\hat{z}$ by $\Theta$ about the axis $(-\sin{\Phi}, \cos{\Phi},0)$ in the $(x,y)$ plane. The parametrisation of the rotation matrix $R$, in terms of the axis and the angle is given by Rodrigues formula \cite{Belongie}
\begin{eqnarray}
    R(\Theta,\Phi)=I+\sin{\Theta} M(\Phi)+ \nonumber \\
    (1-\cos{\Theta})M^2(\Phi), \label{eq:rodrigues_matrix} 
\end{eqnarray}
where $I$ is the identity  matrix and 
\begin{gather}\label{eq: cross product matrix}
    M(\Phi) =\begin{bmatrix}
0 & 0 & \cos{\Phi}\\
0 & 0 & \sin \Phi \\
-\cos \Phi & -\sin \Phi& 0
\end{bmatrix}	
\end{gather}

The rotated configuration (\ref{eq:R2}) depends on the five parameters $\xi, x_s, y_s, \Theta, \Phi$ and in the next section we construct equations describing their time evolution in response to switching the external electric field ${\mathbf{on}}$ and ${\mathbf{off}}$.

\subsection{Frank-Oseen free energy and a dissipation rate}
We adopt the one elastic constant $K_{11} = K_{22} = K_{33} =K$ approximation for the Frank-Oseen free energy per unit length:
\begin{eqnarray}
    \label{eq:free_energy}
    F=\int  \biggl [ \left( \nabla \cdot \nvec\right)^2+\left( \nabla \cross \nvec\right)^2 +2\pi\, \nvec\cdot\nabla \cross \nvec - \nonumber \\
    \frac{\varepsilon_0 \Delta\varepsilon}{2}\frac{p^2}{K} (\Evec \cdot \nvec)^2 \bigg ] d^2r.
\end{eqnarray}
The integral is over the whole plane $\mathbb{R}^2$, and the free energy is written in a dimensionless form, with the cholesteric pitch $p$ as the unit length, and
$Kp$ the energy units. $\varepsilon_0$ is the vacuum permittivity, and the effective electric field  $\Evec = (0,-E,W)^T$, where $E$ is the amplitude of the real electric field applied in the $-y$ direction, and $W$ accounts for effects of the homeotropic anchoring conditions at the cell surfaces of a three-dimensional system. In what follows we set $\sqrt{2K/\varepsilon_0 \Delta\varepsilon p^2}$ as the unit of the electric field, and will use the same notation $E$ and $W$ for the dimensionless electric field and the effective anchoring, respectively. We also introduce the ratio $m$ of the electric and elastic energy in the following way 
\begin{equation} \label{eq: m squared def}
    m^2= (E^2+W^2),
\end{equation}
recall that $E$ and $W$ are given in units of $\sqrt{2K/\varepsilon_0 \Delta\varepsilon p^2}$. As we show below, $m$ controls the relaxation time of the director when the electric field is turn ${\mathbf{on}}$ and ${\mathbf{off}}$. 

The Rayleigh dissipation function per unit length, due to the director reorientation has the form \cite{Long2021}:
\begin{align}
    D &=\frac{1}{2} \int \norm{\frac{d\nvec}{dt}}^2 d^2r = \nonumber \\
     & \frac{1}{2} \int \biggl|\biggl|\frac{\partial \nvec}{\partial \Theta}\dot{\Theta}+\frac{\partial \nvec}{\partial \Phi}\dot{\Phi} + 
     \frac{\partial \nvec}{\partial \rvec_s} \cdot \dot{\rvec_s}+\frac{\partial \nvec}{\partial \xi}\dot{\xi}\biggr|\biggr|^2 d^2r.
    \label{eq:dissipation}
\end{align}
where dots above the parameters indicate time derivatives. $D$ is written in a dimensionless form with $\gamma p^2 /K$ set as the unit of time where $\gamma$ is the rotational viscosity, and the double vertical lines under the integral means the absolute value of the enclosed vector.
The dynamic equations for the skyrmion coarse grained parameters are obtained from the force balance \cite{Long2021}
\begin{gather}\label{eq:balance1}
    \frac{\partial D}{\partial \dot{\Theta}} + \frac{\partial F}{\partial \Theta} = 0, \\ \label{eq:balance2}
    \frac{\partial D}{\partial \dot{\Phi}} + \frac{\partial F}{\partial \Phi} = 0, \\  \label{eq:balance3}
    \frac{\partial D}{\partial\dot{\xi}} + \frac{\partial F}{\partial \xi} = 0, \\ 
  \label{eq:balance4}
    \frac{\partial D}{\partial \dot{x}_s}  = 0, \\ \label{eq:balance5}
    \frac{\partial D}{\partial \dot{y}_s} = 0, 
\end{gather}
here the two last equations take into account that the free energy $F$ does not depend on the skyrmion position $(x_s, y_s)^T$. We fix the reference frame such that $\Evec$ changes in the $(y,z)$ plane only, which will lead to the skyrmion motion along the $x$ axis. Additionally, $\Phi$ after the initial transient regime will stay constant, which we set to $-\pi/2$. Therefore, in the following we consider only Eqs.~(\ref{eq:balance1}), (\ref{eq:balance3}) and (\ref{eq:balance4}). 

This system of equations is not amenable to analytical solution, and thus we developed a  hybrid analytico-numerical method to analyze them. The general algebraic structure of the terms in the dissipation rate $D$ and the free energy $F$ may be written as a product of two tensors
\begin{equation}
  B_{ij}(\Theta) \int T_{ij}(\xi) d^2r,   
\end{equation}
where the summation over repeated indices is implied. Tensors $B_{ij}$ depend on the far field background parameter and are expressed as products $R_{ki}R_{kj}$ or $R_{ki}\partial R_{kj}/\partial \Theta$. The second tensors $T_{ji}(\xi)$ depend only on the thickness $\xi$ of the twist wall and have the form of products of pairs selected from the components of $\nvec^a$ in (\ref{eq:ansatz n0}), spatial derivatives of $\nvec^a$ or its derivatives with respect to $x_s$ and $\xi$. The non-zero components $T_{ij}(\xi)$ are calculated by numerical integration and in all the cases the numerical results can be fitted by simple functions of $\xi$. \textcolor{black}{We find that functions of the form $a + b \xi^n$, where $n=-2,-1,0,1$, can fit well all cases.}


We also assume that the dynamics of $\Theta$ is not dependent on the presence of the skyrmion, and evaluate the lhs and the rhs in Eqs.~(\ref{eq:balance1}) inserting the uniform far field $\nvec_{0}$ into Eqs.~(\ref{eq:free_energy}) and (\ref{eq:dissipation}). As a result we arrive at the following system of dynamical equations:
\begin{eqnarray} 
 \dot{\Theta} = (E \sin \Theta + W \cos \Theta) 
 (E \cos \Theta - W \sin \Theta), \label{eq:theta_ponto}\\
\dot{\xi}  = -d_1(\xi) \biggl ( - \frac{15.5}{\xi^2} + d_2(\xi) (1+ \cos \Theta ) + \nonumber \\
d_3(\xi)\Bigl( W^2 - E^2 + 3 \cos(2\Theta) (E^2 + W^2) + \nonumber \\
 6 EW \sin(2\Theta) \Bigr) \biggr ), \label{eq:xi_ponto} \\ 
\dot{x_s}  = d_4(\xi)\dot{\Theta}.
\label{eq:xs_ponto}
\end{eqnarray} 
Functions $d_1(\xi)$ up to $d_4(\xi)$ are obtained by fitting the numerically calculated $T_{ij}(\xi)$ to simple analytical functions specified above, which renders the following expressions
\begin{eqnarray}
    d_1(\xi)&\approx& \frac{\xi}{16.109 \xi+6.505}\\
    d_2(\xi) &\approx&4\pi \left(\frac{0.179}{\xi}-6.610\right) \\
    d_3(\xi) &\approx&4 (0.544+1.912\xi) \\
    d_4(\xi) &\approx&\frac{4.918+8.535\xi}{15.510+9.714/\xi} \approx 0.551 \xi
    \label{eq:d4}
\end{eqnarray}

\noindent where the function $d_4(\xi)$ is well approximated by a linear relation with $\xi$.

\section{Linear stability analysis}
In this section we carry out a linear stability analysis of the system of Eqs.~(\ref{eq:theta_ponto})-(\ref{eq:xs_ponto}) around the fixed points. The system will evolve towards stable fixed points, and also spend time near them for the case of time-dependent electric fields. The behaviour resulting from the linearised equations sheds light on the evolution of the full non-linear system. Since the dynamics of 
$x_s$ is slaved to that of $\Theta$, Eq.~(\ref{eq:xi_ponto}), it is sufficient to consider only Eqs.~(\ref{eq:theta_ponto}) and (\ref{eq:xi_ponto}).

\begin{figure}[th]
\center
\includegraphics[width=0.98\linewidth]{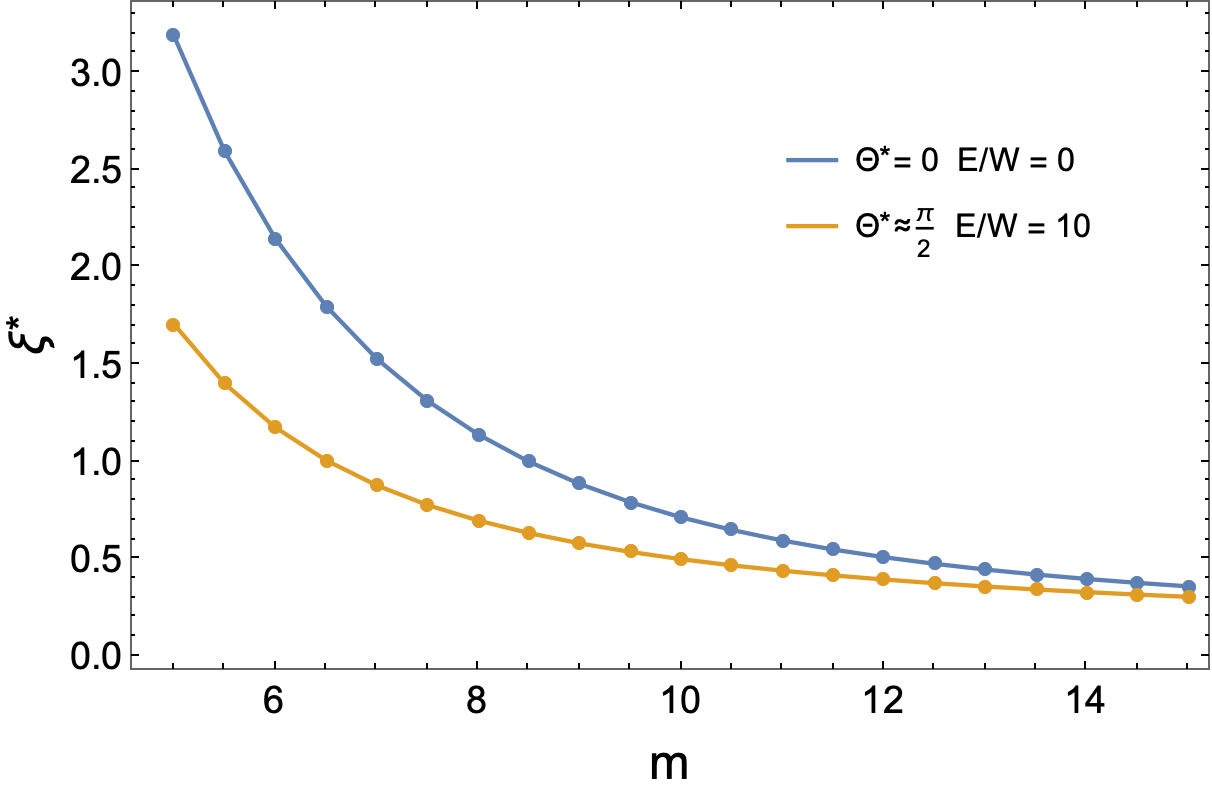}
\caption{{\bf Equilibrium thickness of the skyrmion twist wall}. The stable wall thickness $\xi^*$ as a function of $m=\sqrt{E^2+W^2}$, at two values of the equilibrium $\Theta^*$. }
\label{fig:stablesize}
\end{figure}

Setting the rhs of Eq.~(\ref{eq:theta_ponto}) to zero, renders two fixed point solutions for the polar angle ${\Theta^*}= \tan^{-1} (E/W)$ and ${\Theta^*}= -\tan^{-1} (W/E)$. It turns out that the second solution corresponds to an unstable fixed point. The fixed point for the twist wall thickness $\xi^*$ follows from setting $\Theta=\Theta^*$ on the rhs of Eq.~(\ref{eq:xi_ponto}) and finding the roots of the resulting equation numerically. The resulting fixed point solution $\xi^*$ is plotted in Fig.~\ref{fig:stablesize} as a function of $m$, and is well approximated by $m^{-2}$ in agreement with \cite{Duzgun2018}. In experiments as well as numerical simulations, the skyrmion is only stable in a finite range of electric fields \cite{Duzgun2018} \cite{Tai2020}. 

\begin{figure}[th]
\center 
\includegraphics[width=0.98\linewidth,trim={0.1cm 0.2cm 0.1cm 0.1cm},clip]{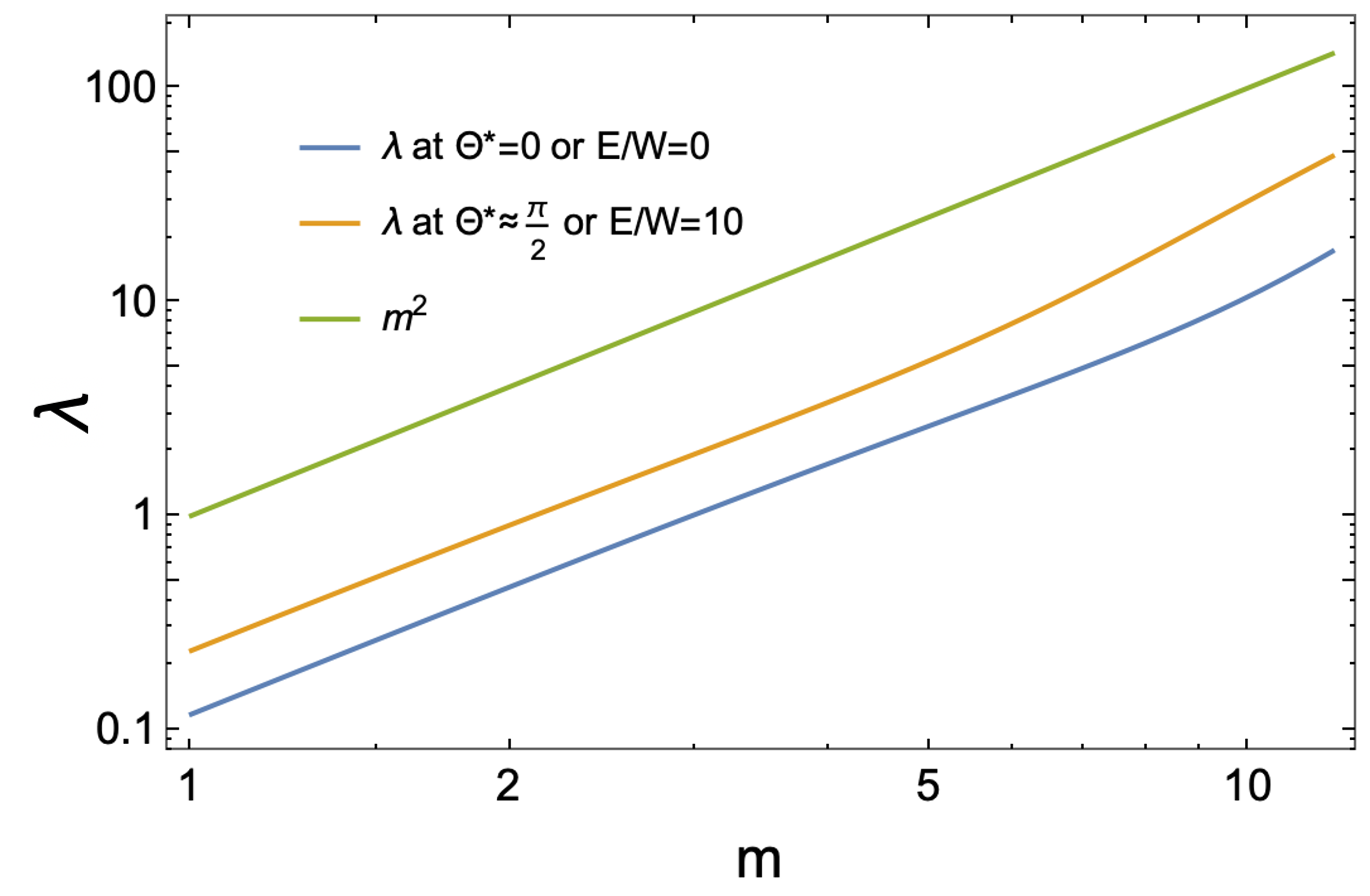}
\caption{{\bf Inverse relaxation time of the twist wall thickness}. $\lambda$, which is the derivative of the rhs of equation (\ref{eq:xi_ponto}) with respect to $\xi$ evaluated at $\xi^*$, as a function of $m=\sqrt{E^2+W^2}$, for two equilibrium values of $\Theta=0$ and $\pi/2$. For comparison the curve $\propto m^2$ is also shown.}
\label{fig:lambda}
\end{figure}

We linearise equations (\ref{eq:theta_ponto}) and (\ref{eq:xi_ponto}) near the fixed point $(\Theta^*, \xi^*)$ 
\begin{gather}
\dot{\Theta} \approx - (E^2+W^2) (\Theta-\Theta^*), \label{eq:linearizedthetap} \\
\dot{\xi}\approx -\left[\xi-\xi^*\right]\lambda(m,E/W),
\label{eq:linxipto}
\end{gather}
where $\lambda(m,E/W)$ is the derivative of the rhs of Eq.~(\ref{eq:xi_ponto}) with respect to $\xi$ evaluated at $\xi^*$. $\lambda$ as a function of $m$ is shown in Fig.~\ref{fig:lambda}, and it is approximately $\propto m^2$ in the relevant range of $m$.


Next, we substitute the solutions $\Theta(t)$ and $\xi(t)$ of the linearised equations \ref{eq:linearizedthetap} and \ref{eq:linxipto} into \ref{eq:xs_ponto}, which renders 
\begin{gather}\label{eq:velocitynearstability}
\dot{x}_s(t)= -m^2e^{-m^2t}(\Theta_{0}-\Theta^*)(\xi^*+\left (\xi_0-\xi^*\right)e^{-\lambda t}),
\end{gather}
where we replaced $d_2(\xi)$ with $\xi$ for simplicity, and $\Theta_0$ is the initial value of the far field polar angle. An approximate equation (\ref{eq:velocitynearstability}) describes the skyrmion velocity in the vicinity of the fixed point $(\Theta^*, \xi^*)$, therefore the initial values $\Theta_{0}$ and $\xi_0$ must be taken close to the fixed point. 

We find that in the relevant range of $m$ $\lambda > 0$, as shown in Fig.~\ref{fig:lambda}, and thus the late time behaviour of the velocity in (\ref{eq:velocitynearstability}) is governed by the term $\propto \exp(-m^2t)$, and the term $\propto\exp(-\lambda t)$ may be neglected. With this in mind, the approximate solution $x_s(t)$ of (\ref{eq:velocitynearstability}) may be written in the following form
\begin{gather}
x_s(t)=\exp(-m^2t) (\Theta_{0}-\Theta^*)\xi^* + x_s(\infty) 
\label{eq:displacement}
\\
\label{eq:deltaxaprox}
\Delta x_s \equiv x_s(\infty) - x_s(0)= \xi^*(\Theta^* - \Theta_{0}),
\end{gather} 
where we have also defined the net skyrmion displacement $\Delta x_s$. 

Equation~\ref{eq:deltaxaprox} highlights two important features in determining the net skyrmion displacement: i) a rotation from the vertical configuration $\Theta_{0}=0$ to a larger $\Theta^*$, e.g when the electric field $E$ is applied, yields a positive displacement increasing with $\Delta\Theta\equiv\Theta^*-\Theta_{0}$; ii) the displacement depends on the stable wall thickness $\xi^*$. Figure~\ref{fig:stablesize} shows that $\xi^*$ decreases with increasing both $m$ and $E/W$. 

Imagine that the electric field $E$ undergoes step-like oscillations between zero and a sufficiently high value, such that $\Theta^*(E)\approx \pi/2$. Additionally, assume that the oscillation period is large enough such that $\Theta$ oscillates between the two equilibrium values $0$ and $\pi/2$. Then, during the field $\mathbf{on}$ state, the net skyrmion displacement $\Delta x_s[\mathbf{off}\rightarrow \mathbf{on}]\approx \pi \xi^*(m_{\mathbf{on}},E)/2$, while the net skyrmion displacement during the $\mathbf{off}$ state is $\Delta x_s[\mathbf{on}\rightarrow \mathbf{off}]\approx -\pi \xi^*(m_{\mathbf{off}},0)$. The expressions in the square brackets indicate the situations when the electric field has changed from 0 to $E$, $[\mathbf{off}\rightarrow \mathbf{on}]$, or from $E$ to 0, $[\mathbf{on}\rightarrow \mathbf{off}]$, also $m_{\mathbf{off}} = W< m_{\mathbf{on}} =  \sqrt{W^2 + E^2}$. Based on the numerical results displayed in Fig.~\ref{fig:stablesize}, we conclude that $\xi^*(m_{\mathbf{off}},0) > \xi^*(m_{\mathbf{on}},E)$ and the net skyrmion's displacement over one period of the electric field is $\Delta x_s[\mathbf{off}\rightarrow \mathbf{on}] + \Delta x_s[\mathbf{on}\rightarrow \mathbf{off}] < 0$ - the skyrmion translates in the negative $x-$direction.

We will show in the next section that the skyrmion velocity can change its sign in the framework of the full non-linear model Eqs.~(\ref{eq:theta_ponto})-(\ref{eq:xs_ponto}).

\section{Full non-linear dynamics}
\subsection{Skyrmion relaxation upon abrupt changes of the electric field }

\begin{figure}[th]
\center
\includegraphics[width=0.98\linewidth]{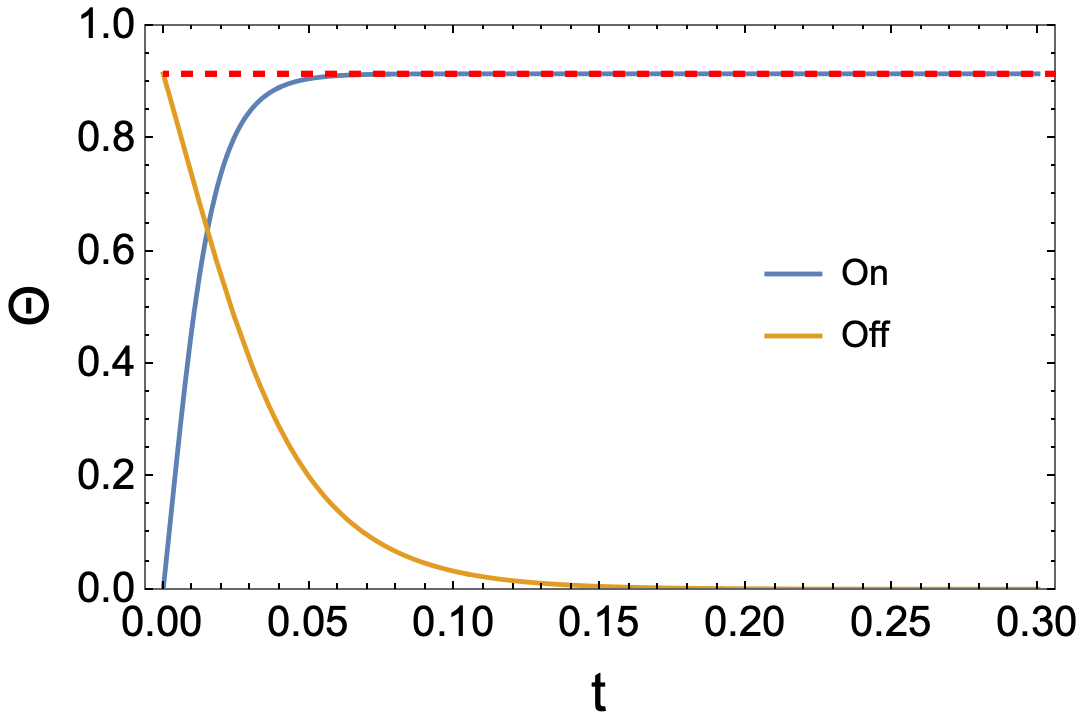}
\caption{{\bf Relaxation dynamics of the polar angle of the far field director}. $\Theta$ as a function of time when the electric field changes abruptly from $0$ to $E$, blue curve. After the angle completely relaxes towards the horizontal dashed red line, the field is turned off, orange curve.  The magnitude of the electric field is 30\% larger than that of the anchoring. $E/W = 0.3$ and $m=\sqrt{E^2 + W^2} = 6$. The horizontal dashed red line corresponds to $\arctan(1.3)$. Time is in units of $\gamma p^2/K$.}
\label{fig:theta_vs_t_full_model}
\end{figure}

We now solve full non-linear equations~(\ref{eq:theta_ponto})-(\ref{eq:xs_ponto}) numerically using {\it Wolfram Mathematica}. We start with the far field director polar angle $\Theta$, which is decoupled from the other two variables. Next we solve for the twist wall thickness $\xi$, which is coupled only to $\Theta$, and finally we determine the skyrmion position $x_s$. Figure~\ref{fig:theta_vs_t_full_model} depicts the time evolution of $\Theta$ when the electric field changes abruptly from $0$ to some value $E$ (blue curve), and then after $\Theta$ reaches the equilibrium value $\arctan(E/w)$, the field is turned off abruptly and $\Theta$ evolves to $0$ (orange curve). The numerical solutions converge quickly to those of the linearised Eq.~\ref{eq:linearizedthetap}, i.e. $\propto \exp(-m^2t)$. If $\xi$ were constant, the skyrmion displacement would look exactly like $\Theta(t)$, see Eq.~(\ref{eq:xs_ponto}).

\begin{figure}[th]
\center
\includegraphics[width=0.9\linewidth]{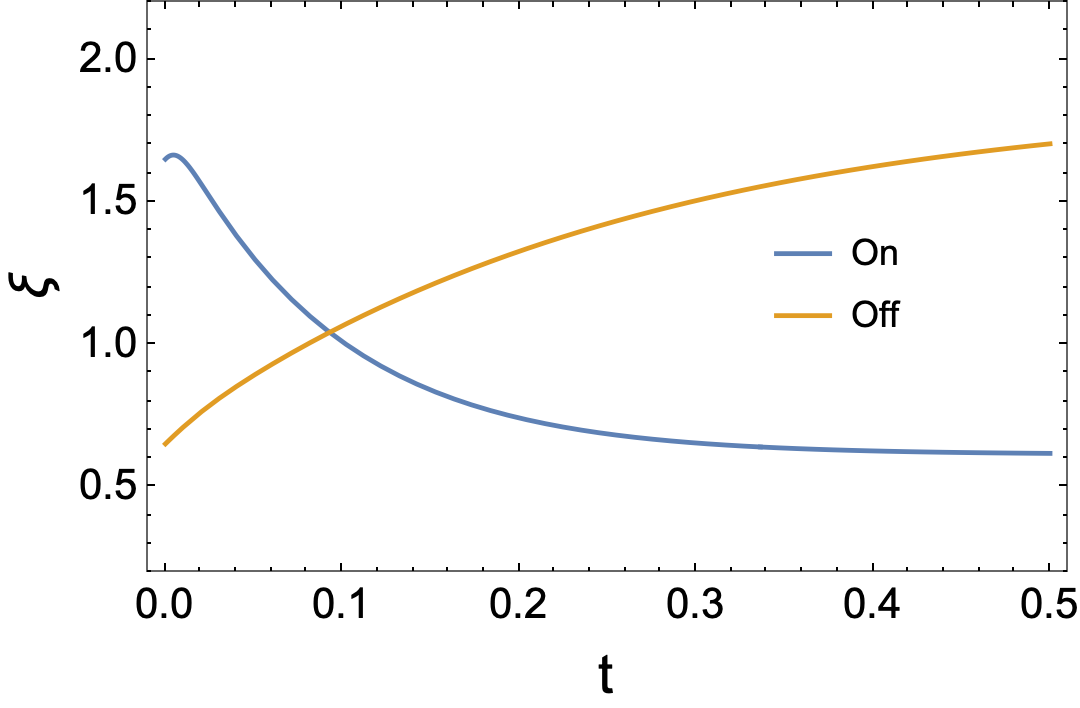}
\caption{{\bf Relaxation dynamics of the thickness of the skyrmion twist wall}. The thickness $\xi$ of the twist wall around the skyrmion core as a function of time, calculated numerically by using Eqs.~(\ref{eq:xi_ponto}) and (\ref{eq:theta_ponto}). The blue curve corresponds to the case when the field is changed abruptly from $0$ to $E$. When $\xi$ reaches the equilibrium  $\xi^*$ the field is turned off. The subsequent dynamics is represented by the orange curve. $m=10$, $E/W=1.2$ and time is in units of $\gamma p^2 /K$.}
\label{fig:xi_one_cycle}
\end{figure}

The internal degree of freedom of the skyrmion, the twist wall thickness $\xi$ is shown in Fig.~\ref{fig:xi_one_cycle}. The non-monotonic early time $\xi(t)$ on the field $\mathbf{on}$ state is caused by the selected initial value $\xi(t=0)>\xi(t=\infty)\equiv\xi^*$.
According to the results in Fig.~\ref{fig:stablesize}, $\xi^*$ decreases with the increasing $E$, therefore more pronounced non-monotonic behaviour of $\xi(t)$ is expected when the field changes from zero to larger values. Thus, the skyrmion speed which is $\propto \xi(t)$ (see Eq.~(\ref{eq:xs_ponto})) is largest just after the field is turned on.

Having determined $\xi(t)$ and $\Theta(t)$ enables calculating the skyrmion trajectory
\begin{equation} \label{eq:position_time_integral}
    x_s(t) \propto \int_0^t \xi(t') \Dot{\Theta}(t')dt',
\end{equation}
where the proportionally constant is the coefficient $0.551$ in the function $d_4(\xi)$, Eq.~(\ref{eq:d4}).

\begin{figure}[th]
\center
\includegraphics[width=0.98\linewidth]{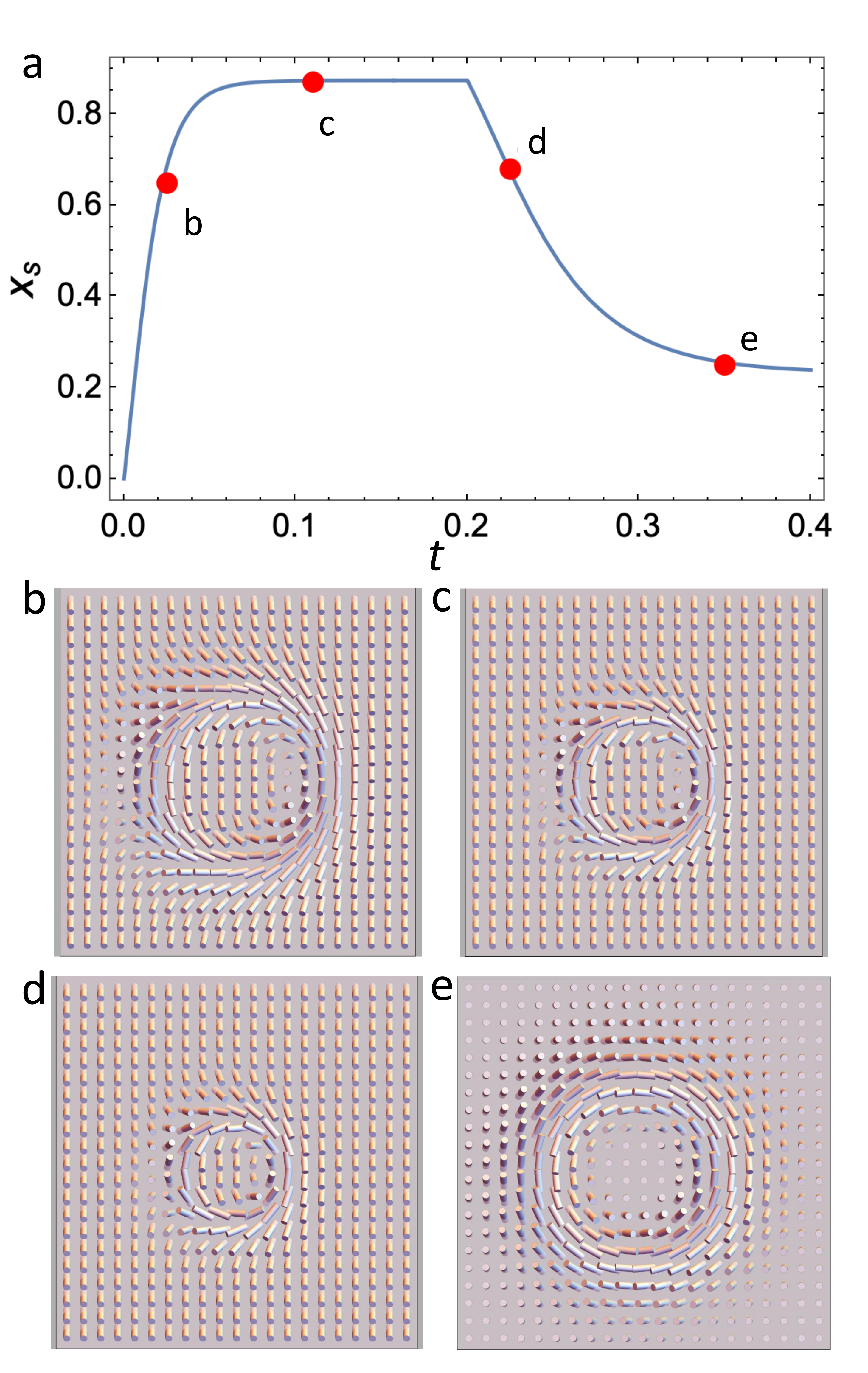}
\caption{{\bf Skyrmion displacement upon turning the field  $\mathbf{on}$  and  $\mathbf{off}$}. (a) Skyrmion position $x_s$ as a function of time $t$ when the electric field is turned on at $t=0$ and then turn off at $t=0.2$. In the $\mathbf{on}$ state $m_{\mathbf{on}}=\sqrt{E^2+W^2}=6$ and $E/W = 1.3$. When the field is turned off the background and the position take longer to relax because $m_{\mathbf{off}} = W <m_{\mathbf{on}}$. Time is in units of $\gamma p^2 / K$. (b)-(e) show director configurations represented by the cylindrical rods. The configurations correspond to the times indicated  by (b)-(e) in panel (a). (b) $\Theta = 0$; (c) $\Theta=\arctan(1.3)$, at this point $\xi=1.2$ is at it lowest value. (d) and (e), $\xi$ increases and the skyrmion becomes axisymmetric again.}
\label{fig:XS_and_confs}
\end{figure}

We evaluate the integral in Eq.~(\ref{eq:position_time_integral}), starting with the axisymmetric skyrmion configuration, Eq.~(\ref{eq:ansatz n0}), (corresponding to the zero-field fixed point). At $t=0$ the electric field is set to a value $E$, and the skyrmion morphs (see Fig.~\ref{fig:XS_and_confs}\textcolor{black}{(b)}) and is allowed to reach the second fixed point, shown in Fig.~\ref{fig:XS_and_confs}\textcolor{black}{(c)}. Then, after equilibrium is reached, the field is turned $\mathbf{off}$ and the skyrmion relaxes back (see Figs.~\ref{fig:XS_and_confs}\textcolor{black}{(d)} and \textcolor{black}{(e)}) to the axisymmetric form. The skyrmion trajectory $x_s(t)$ corresponding to this protocol including several representative skyrmion  configurations are shown in Fig.~{\ref{fig:XS_and_confs}.

When the field is turned $\mathbf{on}$, the far field director $\nvec_{0}$ tilts away from $\hat{z}$, dragging the skyrmion along $\hat{x}$. When the background reaches $\Theta^*=\arctan(E/W)$ the skyrmion stops, with the corresponding configuration shown in Fig.~\ref{fig:XS_and_confs}\textcolor{black}{(c)}. Next the field is turned $\mathbf{off}$ and $\nvec_{0}$ rotates back to $\hat{z}$, Fig.~\ref{fig:XS_and_confs}\textcolor{black}{(e)}. During this field-$\mathbf{off}$ state, the relaxation is $(E^2+W^2)/W^2=2.7$ times slower than in the preceding field-$\mathbf{on}$ state.

Equation~(\ref{eq:deltaxaprox}) is valid in the linear regime and predicts a negative displacement under such protocol, because the stable thickness is larger for the $\mathbf{off}$ state, but the opposite is observed here. This can be understood by analysing Eq.~(\ref{eq:position_time_integral}). Indeed, the rate of change of $\Theta$ is maximal right after the field changes, and is significantly reduced by approaching to the equilibrium where the linear regime holds. Therefore, the initial non-linear transient of $\Theta$ is more relevant in determining the net displacement, than the stable thickness $\xi^*$ controlling the skyrmion displacement at late times. This transient is affected by the initial conditions and field parameters, which will be discussed in the following section.

\subsection{Steady state cycle in pulse width modulated electric fields}
We discuss the skyrmion dynamics in a pulse width modulated electric field. The electric field is activated for a fraction of the period, referred to as the duty cycle, and then turned off. The period $T$ is the sum of the field ${\mathbf{off}}$ and ${\mathbf{on}}$ times.

\subsubsection{Driving protocol 1)}

We first consider a driving protocol where the electric field has only one component in the $-\hat{y}$ direction, denoted by protocol 1). In the course of time the skyrmion trajectory will approach a characteristic closed loop when plotted in the phase plane $(\Theta,\xi)$, with one example shown in Fig.~\ref{fig:close_cycle}. We call such a loop the steady state cycle, in the sense that as long as the field continues to oscillate the system will cycles along the loop indefinitely.

\begin{figure}[th]
\center
\includegraphics[width=0.98\linewidth]{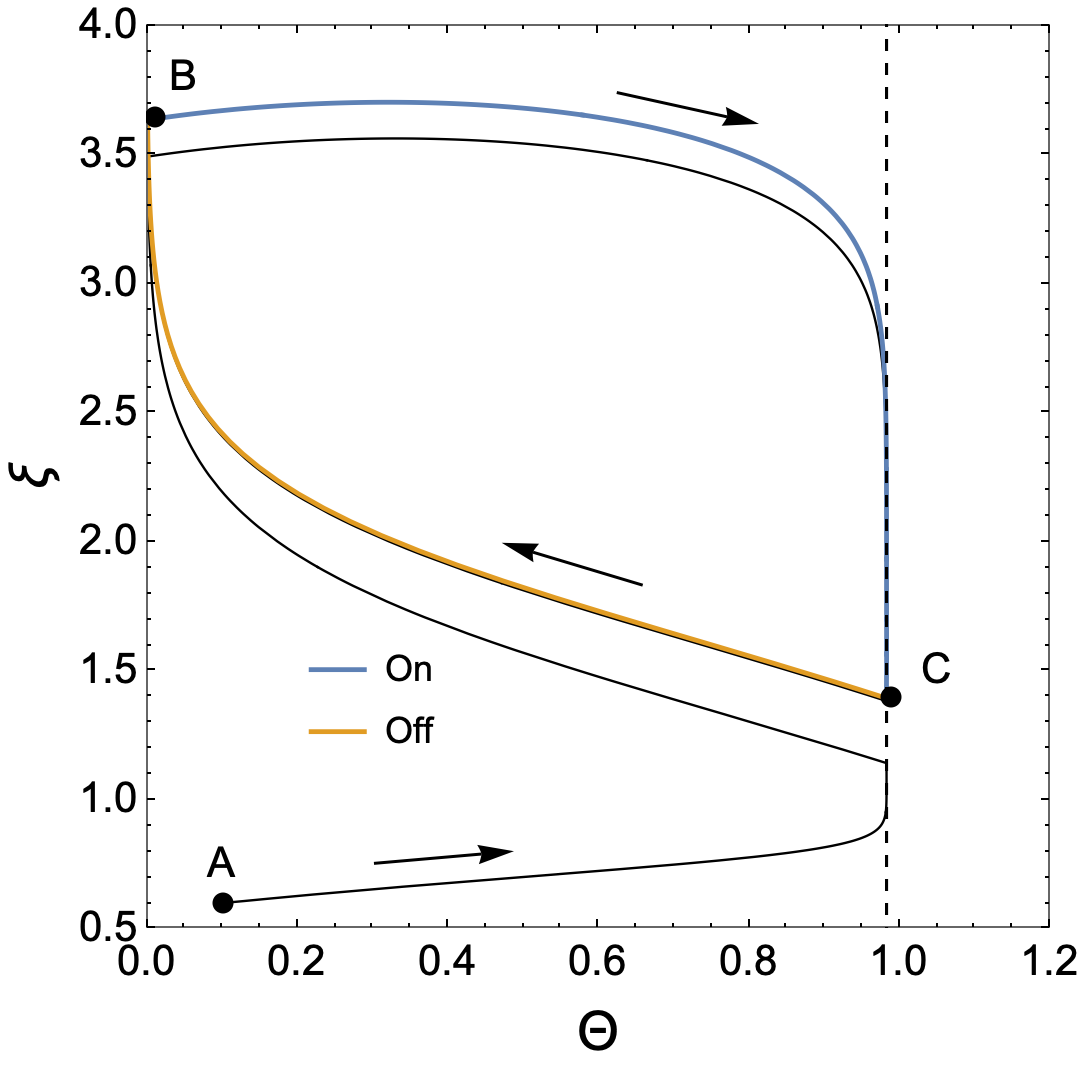}
\caption{{\bf Steady state cycle upon periodic turning the field on and of}. Skyrmion trajectory in the $(\Theta,\xi)$ phase plane upon periodic switching the field $\mathbf{on}$ and $\mathbf{off}$. The initial configuration corresponds to point \textit{A} where the field is $\mathbf{on}$. The transient trajectory is shown by the black line, while the blue (field-$\mathbf{on}$ state) and the orange (field-$\mathbf{off}$ state) lines correspond to the long time steady state cycle. The black arrows indicate the direction of the motion. The field is turned $\mathbf{on}$ at point \textit{B} and $\mathbf{off}$ at \textit{C}. The net skyrmion displacement is proportional to the area enclosed by the loop. The area between the field-$\mathbf{on}$ (blue) trajectory and the horizontal axis is larger than the area between the field-$\mathbf{off}$ (orange) trajectory and the horizontal axis, which results in the skyrmion motion in the $+\hat{x}$ direction. The field parameters are $m=7$, $E/W=1.5$, $T=1/3$ and ${\cal D} = 0.5$.}
\label{fig:close_cycle}
\end{figure}

The repeating cycles yield skyrmion motion with the average velocity $\Delta x_s /T$, where $\Delta x_s$ is the net skyrmion displacement over one period $T$. It is instructive to change the integration variable in Eq.~(\ref{eq:position_time_integral}) from time $t'$ to $\Theta$ giving the net skyrmion spatial displacement along the steady state cycle 
 \begin{equation} \label{eq:position_angle_integral}
    \Delta x_s \propto \left( \int_{\Theta_A}^{\Theta_B} d\Theta \xi \left( t_{\mathbf{on}}^{-1}(\Theta)\right )- \int_{\Theta_A}^{\Theta_B} d\theta \xi \left (t_{\mathbf{off}}^{-1}(\Theta) \right ) d\Theta \right ),
\end{equation}
where the first integral is taken over the $\mathbf{on}$ branch (the blue curve in Fig.~\ref{fig:close_cycle} ) of the cycle, and the second integral over the $\mathbf{off}$ branch (orange curve in Fig.~\ref{fig:close_cycle}) of the cycle. $\Theta_A$ and $\Theta_B$ correspond to the steady state values of the $\Theta$ at times when the field undergoes jumps from $0$ to $E$ and then back to $0$, respectively. Equation \ref{eq:position_angle_integral} shows that the steady state net skyrmion displacement is proportional to the area enclosed by the steady cycle loop. The skyrmion trajectory $x_s(t)$ corresponding to  Fig.~\ref{fig:close_cycle} is shown in Fig.~\ref{fig:oscilating_position_1}. The steady state cycle is divided into the $\mathbf{on}$ branch (blue curve in Fig.~\ref{fig:close_cycle}) where the skyrmion moves in the $+\hat{x}$ direction, and the $\mathbf{off}$ (orange curve in Fig.~\ref{fig:close_cycle}) branch where it moves in the $-\hat{x}$ direction. The velocity is related to the rate of change of $\Theta$. In the linear stability analysis, we obtained $\Theta\propto\exp(-m^2 t)$. Therefore, the relaxation of $\Theta$ on the $\mathbf{off}$ branch is $1+(E/W)^2$ times slower compared to the relaxation on the $\mathbf{on}$ branch. The larger the electric field compared to the anchoring, the longer the director takes to relax to the vertical configuration promoted by the effective anchoring.

\begin{figure}[th]
\center
\includegraphics[width=0.98\linewidth]{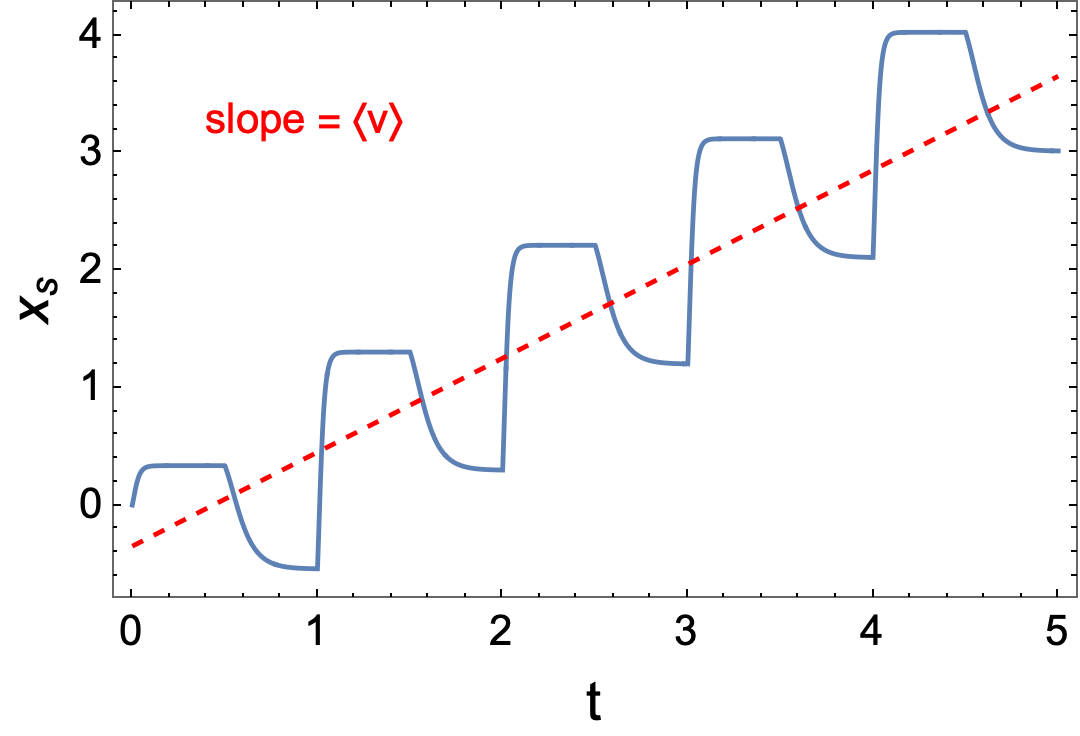}
\caption{{\bf Skyrmion displacement upon periodic turning the field $\mathbf{on}$  and  $\mathbf{off}$}. Skyrmion trajectory $x_s(t)$ corresponding to the situation depicted in Fig.~\ref{fig:close_cycle}. The field parameters are $m=7$, $E/W=1.5$, $T=1/3$ and ${\cal D} = 0.5$.}
\label{fig:oscilating_position_1}
\end{figure}

The mean skyrmion velocity $\langle v \rangle$ is given by the net displacement (the area enclosed by the steady cycle loop) divided by the period $T$. We compute $\langle v \rangle$ as the average slope of the numerical steady cycle trajectories $x_s(t)$. Note that $\langle v \rangle$ equals the slope of the red dashed line in Fig.~\ref{fig:oscilating_position_1}. 
%
\begin{figure}[th]
\center
\includegraphics[width=0.98\linewidth]{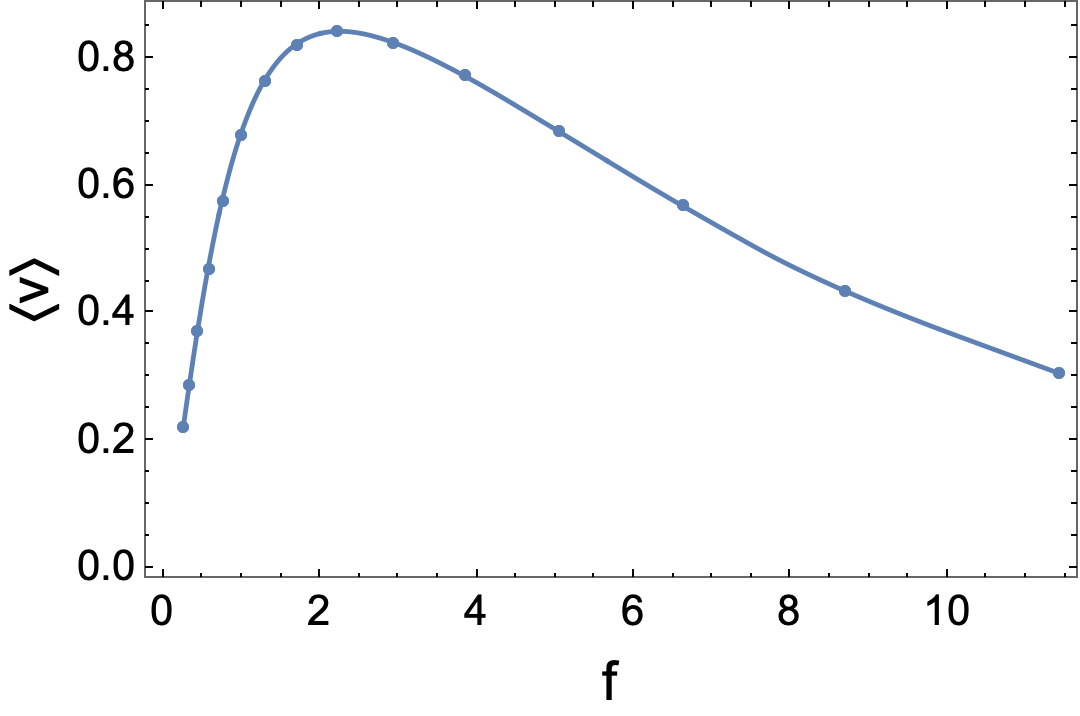}
\caption{{\bf Skyrmion average velocity}. $\langle v \rangle$ of the skyrmion calculated numerically as a function of the frequency $f=1/T$. $m_{on}=\sqrt{E^2 + W^2}=10$, $E/W=1.5$ and ${\cal D}=0.47$.}
\label{fig:velocity_vs_f}
\end{figure}
%
$\langle v \rangle$ as a function of the frequency $f = 1/T$ of the pulsed electric field $E(t)$, at selected values of the duty cycle and the field amplitude is shown in Fig.~\ref{fig:velocity_vs_f}. $\langle v \rangle$ exhibits a maximum at a frequency $f_{max}$ and is always positive, which is specific to the selected profile of the effective electric field, which also includes the effective anchoring $W$ as its $z$ component. As we show below, velocity reversal may be achieved by introducing a pulsed $z-$component of the electric field $E_z(t)$, such that $(W + E_z)^2 + E^2$ remains constant. 

To better understand the emergence of the maximum in $\langle v \rangle$ we examine how steady state cycles are affected by changing the frequency, which is shown in Fig.~\ref{fig:loops_prot1}. $\Theta$ increases when the field is turned $\mathbf{on}$, and decreases when it is turned $\mathbf{off}$. Decreasing the frequency from $f_{max}$ allows $\Theta$ to relax completely. As a result, the steady state cycle shown in Fig.~\ref{fig:loops_prot1}(a), remains unchanged by reducing the frequency even further. The system spends most of the time at the endpoints of the branches. Unchanging steady state cycles lead to a net displacement independent on $f$. A constant displacement divided by the growing period, results in a linear decrease in $\langle v \rangle$ in the low-frequency regime of Fig.~\ref{fig:velocity_vs_f}.

\begin{figure}[th]
\center
\includegraphics[width=\linewidth]{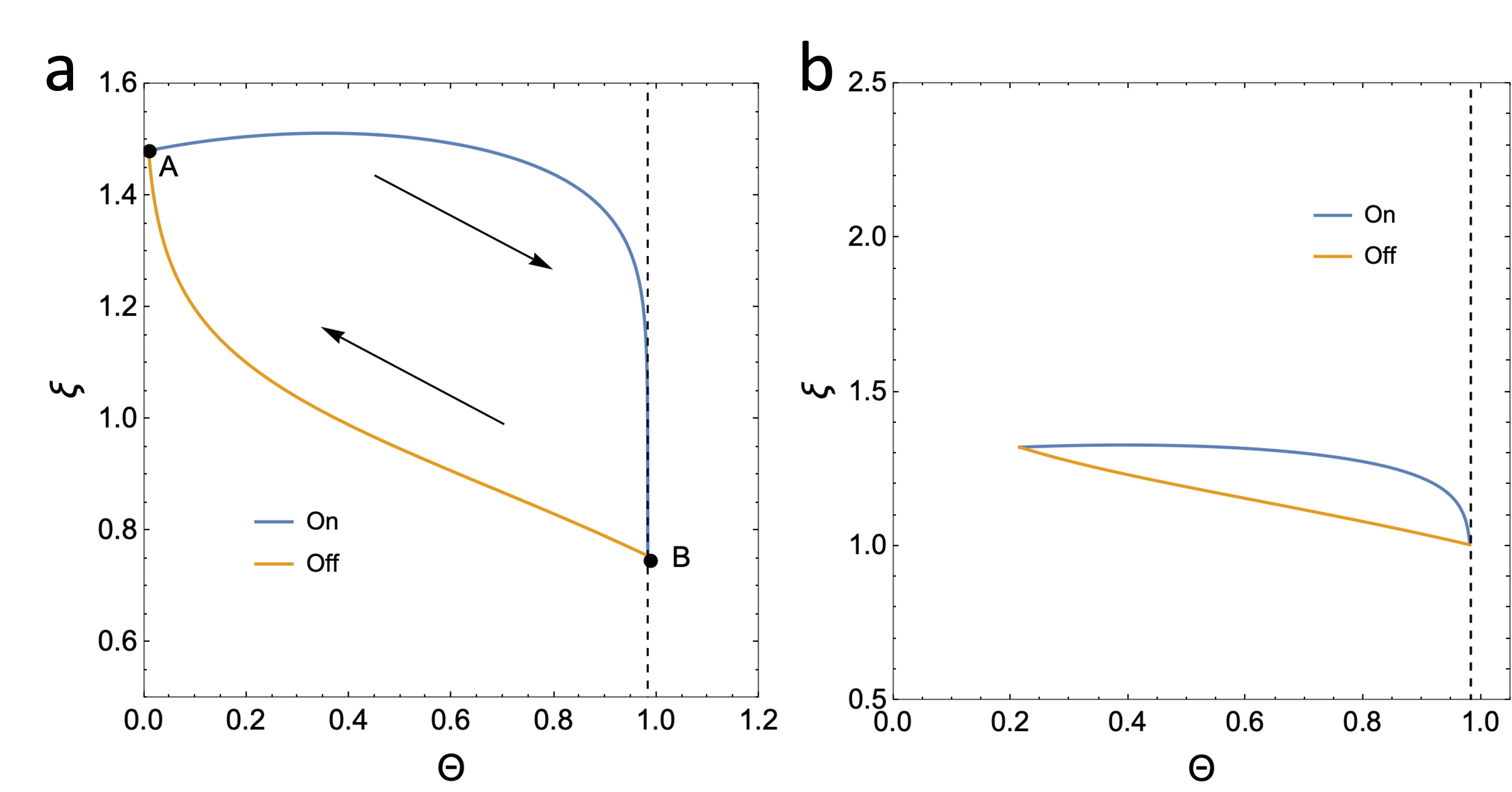}
\caption{{\bf Steady state cycle shrinks as the frequency increases}. Steady state cycles of the system in the $(\Theta,\xi)$ configuration space. The dashed vertical lines marks the stable $\Theta$ when the field is turned on. The values of the parameters are $m_{\mathbf{on}}=\sqrt{E^2+W^2}=10$, $E/W=1.5$, ${\cal D}=0.5$, (a) $f=1$, and (b) $f=8$. Low frequency in (a) allows the system to cover the full angular range. There is also a large difference in $\xi$ with the electric field on and off, as compared to the high frequency loop in (b). }
\label{fig:loops_prot1}
\end{figure}


The reduction in $\langle v \rangle$ at high frequencies is due to the decrease in the range $\Delta\Theta$ of the variation of the background director tilting, and the reduction of the difference between $\xi_{\mathbf{on}}$ (blue curve in Fig.~\ref{fig:loops_prot1}(b)) and $\xi_{\mathbf{off}}$ (orange curve in Fig.~\ref{fig:loops_prot1}(b)) in the two branches of the steady state cycle. We can write the average skyrmion velocity in the following way

\begin{equation}
   \langle v \rangle \propto f \Delta\Theta(\overline{\xi_{\mathbf{on}}}- \overline{ \xi_{\mathbf{off}}}), 
   \label{eq:mean_velocity}
\end{equation}
where we used Eq.~(\ref{eq:position_angle_integral}) and introduced an average over $\Theta$  width $\overline{\xi_{\mathbf{on}}}$ when the field is $\mathbf{on}$ and  $\overline{\xi_{\mathbf{off}}}$ when the field is $\mathbf{off}$. 
On the other hand $\Delta \Theta < \Dot{\Theta}_{max}/f$, and the maximal angular velocity $\Dot{\Theta}_{max}$ depends only on the field strength, and not on $f$, so increasing the frequency will decrease the angular displacement $\Delta\Theta$. This renders $\langle v \rangle < \Dot{\Theta}_{max} (\overline{\xi_{\mathbf{on}}}-\overline{\xi_{\mathbf{off}}})$ which tends to zero as $f$ increases due to the decrease of the difference in brackets.

\begin{figure}[th]
\center
\includegraphics[width=0.99\linewidth]{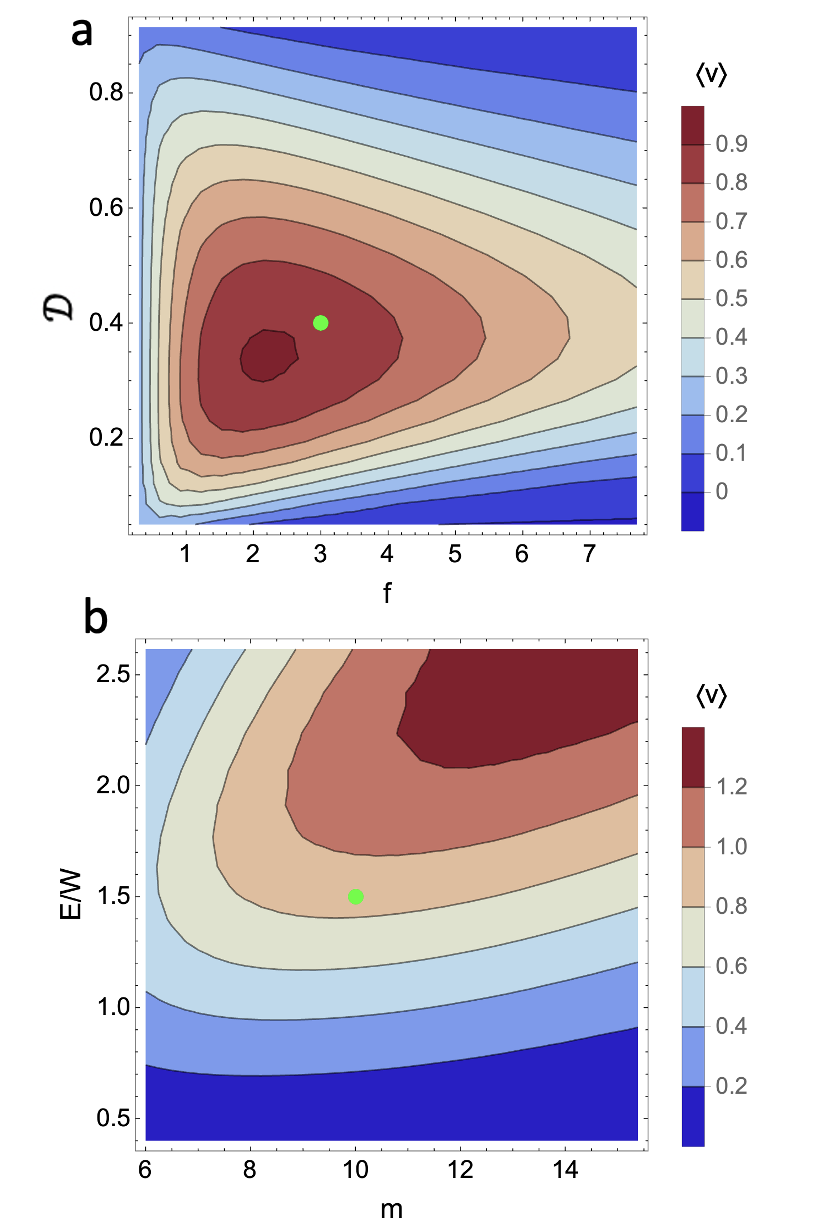}
\caption{{\bf Heat map of the skyrmion average velocity for driving protocol 1)}. Colour coded average velocity $\langle v \rangle$ of skyrmions moving under pulse width modulated electric fields. (a) $\langle v \rangle$ as a function of ${\cal D}$ and $f$ at $m=10$ and $E/W=1.5$ corresponding to the green circle in (b). (b) $\langle v \rangle$ as a function of $m$ and $E/W$ at $f=3$ and ${\cal D}=0.4$ corresponding to the green circle in (a). }
\label{fig:velocity_maps_prot1}
\end{figure}

Figure~\ref{fig:velocity_maps_prot1} summarises the effects of the field strength, the frequency and the duty cycle upon the skyrmion average velocity. The velocity is positive in the studied region of the parameter space. This result is directly related to the relaxation behavior of $\xi(t)$, shown in Fig.~\ref{fig:xi_one_cycle}, which in this regime of parameters is always monotonically decreasing/increasing along the $\mathbf{on}$/$\mathbf{off}$ states. Consequently the difference $(\overline{\xi_{\mathbf{on}}}-\overline{\xi_{\mathbf{off}}}) > 0$ for all steady cycles, and the skyrmion velocity is positive.
This analysis suggests that for velocity reversal there must be an extended range of a markedly non monotonic behaviour of $\xi(t)$, such that it is also accompanied by a significant $\Delta\Theta$. We observe a weak non monotonicity in $\xi_{\mathbf{on}}(t)$ (blue curve in Fig.~\ref{fig:xi_one_cycle}) at early times, which in principle may give rise to velocity reversal. However, the corresponding range of the model parameters around that bump on $\xi_{\mathbf{on}}(t)$ is such that the calculated skyrmion speeds are $\sim 10^{-5}$ and our numerical method is not capable to resolve eventual changes in the direction of motion.

\begin{figure}[th]
\center
\includegraphics[width=0.98\linewidth]{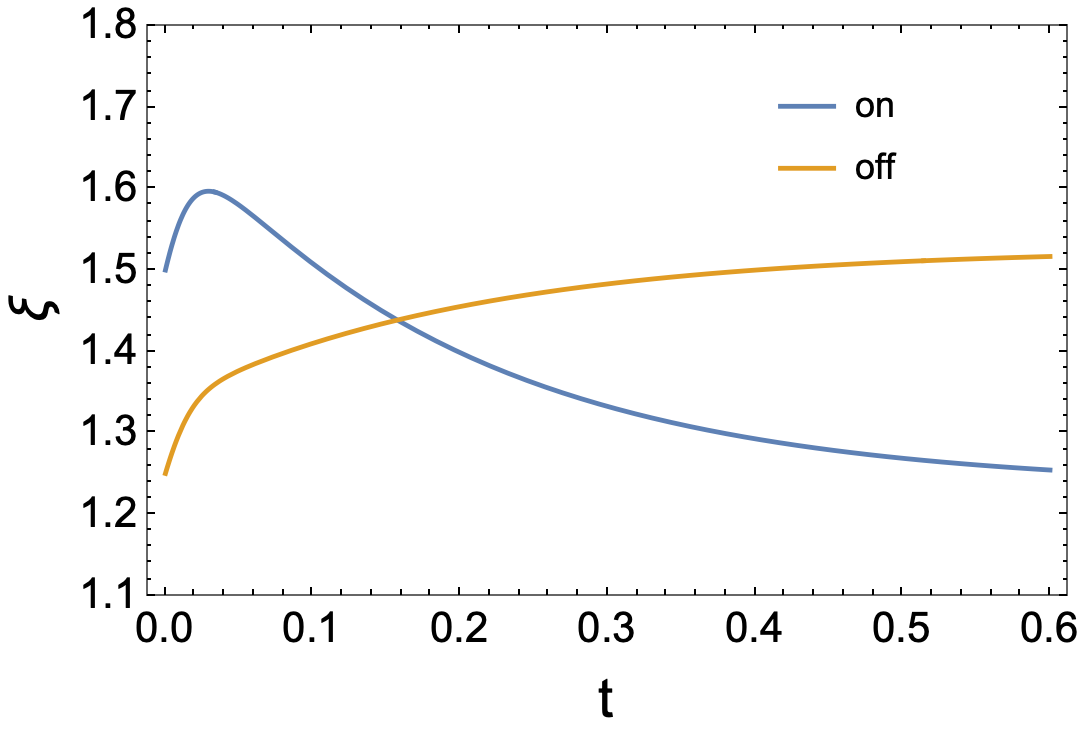}
\caption{{\bf Relaxation dynamics of the twist wall thickness for driving protocol 2)}. The thickness $\xi$ as a function of time when the electric field $(0, -E, E_z)$ is turned ${\mathbf{on}}$ at $t=0$, blue curve. After $\xi$ relaxes completely, the electric field is turned ${\mathbf{off}}$, and the corresponding $\xi(t)$ is shown by the orange curve, where $t=0$ coincides with the time of turning the field ${\mathbf{off}}$. For both curves $m=\sqrt{E(t)^2 +(W+E_z(t))^2}=7$. The units of time is $\gamma p^2 /K$. }
\label{fig:xi_one_cycle_prot2}
\end{figure}

Below we show that it is possible to fix this issue by resorting to another driving protocol. Indeed, in experiments \cite{Ackerman2017} and \cite{Sohn2018}, the background angle $\Theta$ relaxes faster towards zero when the field is turned ${\mathbf{off}}$. This is incompatible with the driving protocol discussed above, because the ${\mathbf{off}}$ relaxation time $\tau_{\mathbf{off}}$ is $1+(E/W)^2$ times smaller, when compared to the ${\mathbf{on}}$ relaxation time $\tau_{\mathbf{on}}$. 

 \subsubsection{Driving protocol 2)}

 We proceed to consider a driving protocol which renders $\tau_{\mathbf{on}}=\tau_{\mathbf{off}}$, which is sufficient to obtain velocity reversal. To this end we introduce an additional electric field in the $z-$direction $E_z(t)$, such that $m = \sqrt{E(t)^2 +(W+E_z(t))^2}=const$. 

An example of the relaxation of $\xi$ under this protocol is shown in Fig.~\ref{fig:xi_one_cycle_prot2}. The first difference, when compared to protocol 1) discussed in Fig.~\ref{fig:xi_one_cycle}, is the reduced difference between $\xi_{\mathbf{on}}(t)$ (blue curve) and $\xi_{\mathbf{off}}(t)$ (orange curve), which will lead to an overall reduction of the skyrmion speed. Secondly, and more importantly, the $\xi_{\mathbf{on}}(t)$ has a significant and extended regime of non-monotonic behaviour at early times, which enables to tune the model parameters, such that  $(\overline{\xi_{\mathbf{on}}}-\overline{\xi_{\mathbf{off}}})$ changes sign. In Fig.~\ref{fig:loops_prot2}(a), which corresponds to low frequencies, the steady state cycle explores the whole range of $\Theta$ and the loop passes through both fixed points. At the same time $\xi$ is larger while the field is $\mathbf{on}$, and the skyrmion moves in $+\hat{x}$ direction. At higher frequency, in Fig.~\ref{fig:loops_prot2}(b), the steady state cycle shrinks and is attached to the $\mathbf{off}$ fixed point ${\Theta}^*=0$.

\begin{figure}[th]
\center
\includegraphics[width=\linewidth]{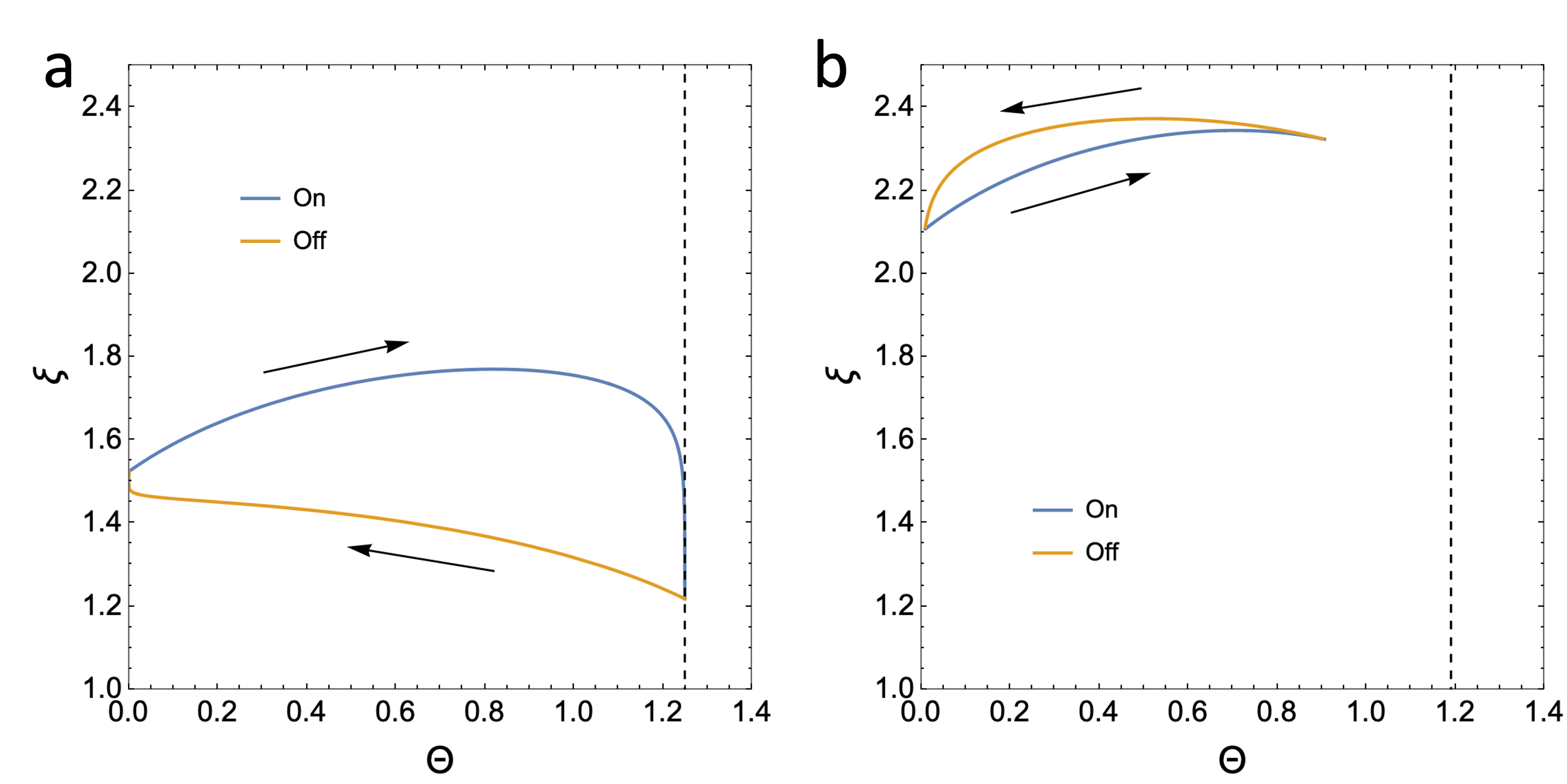}
\caption{{\bf Steady state cycles for driving protocol 2)}. Steady state cycles of the system in the $(\Theta,\xi)$ configuration space for protocol 2), where a vertical component of the electric field $E_z(t)$ is introduced such that $m=\sqrt{E(t)^2+(W+E_z(t))^2}=7$ for both field-${\mathbf{on}}$ and field-${\mathbf{off}}$ states. The dashed vertical lines marks the stable $\Theta$ when the field is ${\mathbf{on}}$. The values of the other parameters are $E/(W+E_z)=1.25$, ${\cal D}=0.3$, (a) $f=1$, and (b) $f=7$. Low frequency in (a) allows the system to cover a large angular range and $(\overline{\xi_{\mathbf{on}}}-\overline{\xi_{\mathbf{off}}}) > 0$ leading to a positive skyrmion velocity. High frequency in (b) reduces the angular range and $(\overline{\xi_{\mathbf{on}}}-\overline{\xi_{\mathbf{off}}}) < 0$, which results in a negative skyrmion velocity. }
\label{fig:loops_prot2}
\end{figure}

Figure \ref{fig:xi_one_cycle_prot2} reveals that when the field $(0,-E,E_z)$ is turned ${\mathbf{on}}$, $\xi$ passes through a broad maximum and relaxes non-monotonically to the new and smaller stable value. If the frequency is large enough and the duty cycle is small enough, the $\xi_{\mathbf{on}}(t)$ branches follow the short time increasing part of the $\mathbf{on}$ curve in Fig.~\ref{fig:xi_one_cycle_prot2}. This behaviour, is depicted in Fig.~\ref{fig:loops_prot2}(b) by the blue curve. Next, after the field is turned ${\mathbf{off}}$ the system follows the orange branch in Fig.~\ref{fig:loops_prot2}(b).
Surprisingly, just after setting the field to zero we observe a growing $\xi_{\mathbf{off}}(t)$, which leads to $(\overline{\xi_{\mathbf{on}}}-\overline{\xi_{\mathbf{off}}}) < 0$
 and the velocity reversal. 

\begin{figure}[th]
\center
\includegraphics[width=0.98\linewidth]{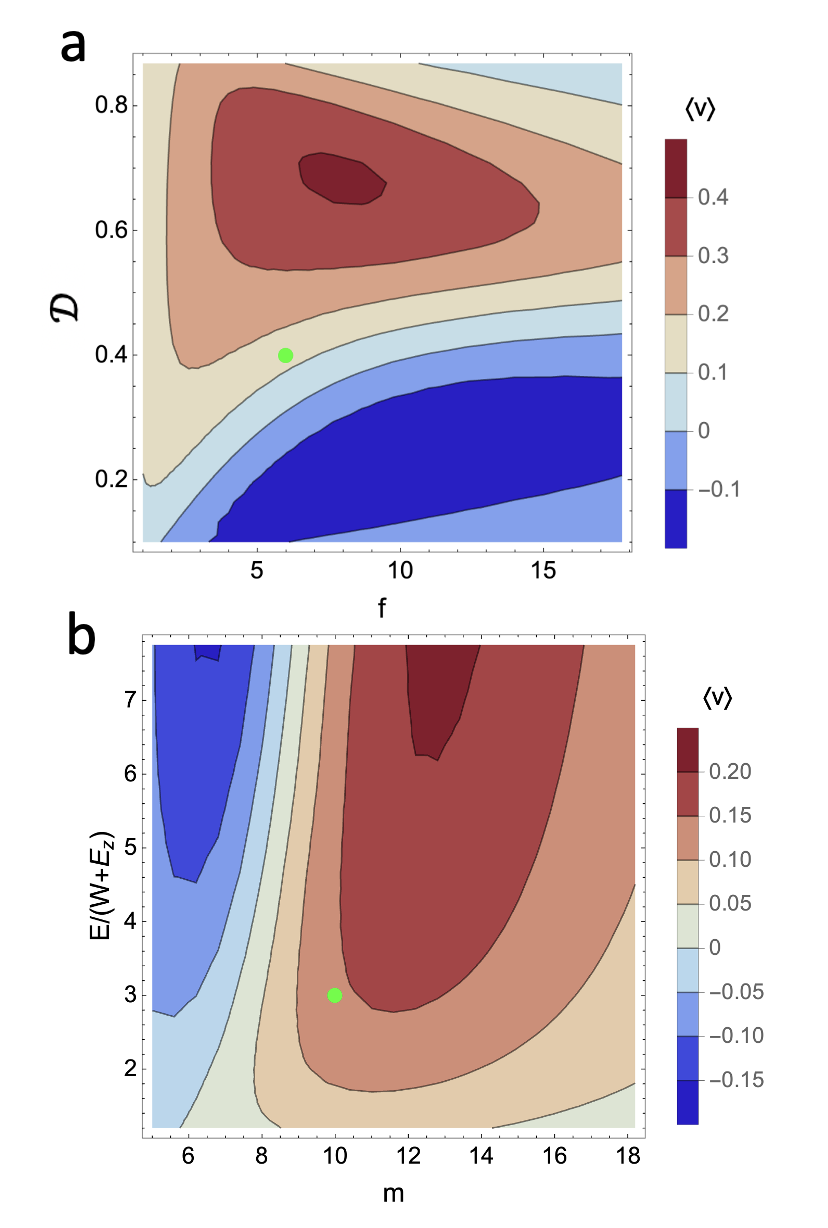}
\caption{{\bf Heat map of the skyrmion average velocity for driving protocol 2)}.Colour coded average velocity $\langle v \rangle$ of skyrmions moving under an electric field $(0,-E(t),W+E_z(t))^T$, where both time-dependent components are synchronously pulsing with the frequency $f$ and the duty cycle ${\cal D}$. (a) $\langle v \rangle$  as a function of ${\cal D}$ and $f$ at $m=10$ and $\arctan(E/(W+E_z))=1.25$, corresponding to the green circle in (b). (b) $\langle v \rangle$ as a function of $m$ and $E/(W+E_z)$ at $f=6$ and ${\cal D}=0.4$, corresponding to the green circle in (a).}
\label{fig:velocity_maps_prot2}
\end{figure}

 The colour coded velocity profile in the $(f,{\cal D})$ and $(m, E/W+E_z)$ planes are shown in Fig.~\ref{fig:velocity_maps_prot2}. Fig.~\ref{fig:velocity_maps_prot2}(b) shows that the skyrmion speed increases with $E/(W+E_z)$, but the direction of motion cannot be reversed by changing this parameter. The most efficient way to reverse the skyrmion velocity is by using the duty cycle. 
 Indeed, by analysing Fig.~\ref{fig:velocity_maps_prot2}(a) we find that the velocity can be reversed by changing $\tau$ for any $f$. By contrast reversing the velocity by changing $f$ can only be achieved in a narrow range of $\tau$. 

\section{Conclusions and Outlook}

We developed a 2D coarse grained model of the driven motion of the LC skyrmions. The starting point was a five parametric axisymmetric {\it Ansazt} for the director field of a skyrmion with skyrmion number equal to one. The five parameters are the polar and azimuthal angles of the far field director; the width of the twist wall around the skyrmion core, and the location of the skyrmion center. Assuming the driving electric field varies in a plane perpendicular to the skyrmion plane, the azimuthal angle can be ignored, as it quickly relaxes to the plane of the electric field, while the skyrmion motion proceeds in the direction normal to that plane. Therefore, only three parameters (degrees of freedom) are relevant for the skyrmion dynamics.

The linear stability analysis of the governing dynamical equations demonstrates that the far field director relaxes with the characteristic time $m^{-2}$, where $m=m_{\mathbf{on}}\equiv\sqrt{E^2 + W^2}$ or $m=m_{\mathbf{off}}\equiv W$ for the field-${\mathbf{on}}$ or field-${\mathbf{off}}$ states, respectively. Therefore, for this driving protocol the far field background relaxes slower towards the fixed point $\hat{z}$, corresponding to zero field. The skyrmion velocity is given by the product of the rate of change of the polar angle $\Theta$ and the fixed point value of the twist wall thickness $\xi^*$ (different for field-${\mathbf{on}}$ and field-${\mathbf{off}}$). The resulting displacement of the skyrmion over one period of the pulse width modulated electric field is given for large periods by $\Delta x_s = \pi ( \xi^*(m_{\mathbf{on}},E) - \xi^*(m_{\mathbf{off}},E))$, which is negative as shown in Fig.~\ref{fig:stablesize}. 

The full non linear model predicts $\Delta x_s \propto \Delta\Theta(\overline{\xi_{\mathbf{on}}}-\overline{\xi_{\mathbf{off}}})$, where $\Delta\Theta$ is the (positive) range covered by the polar angle $\Theta$ over one period, and the symbols with the bars denote $\xi(t)$ averaged over $\Theta$ for the  field-$\mathbf{on}$ and field-$\mathbf{off}$ states. We found that $\Delta x_s$ is always positive for the driving protocol 1), i.e. the average velocity does not change direction and exhibits a single maximum at certain values of the field frequency, duty cycle, and effective field strength.   

Velocity reversal is observed for driving protocol 2) where $m=const$ independent of the value of the electric field. This renders the relaxation times for the far field director equal in the ${\mathbf{on}}$ and ${\mathbf{off}}$ states of the cycle. We observed positive velocity at low frequencies and large duty cycles when the electric field is ${\mathbf{on}}$ most of the time; and negative velocity at high frequencies and small duty cycles when the electric field is ${\mathbf{off}}$ most of the time. 

Here we focused on LCs with positive dielectric anisotropy, and one possible future direction is to study systems with negative one, which is the case of most experimental systems. Another possibility is to investigate the effects of varying the skyrmion core size, which was kept constant equal to the half of cholesteric pitch. Finally, generalisation of the presented approach to many skyrmion systems could shed light on the out-of-equilibrium collective behaviour of skyrmions.

\section*{Acknowledgements}

We acknowledge financial support from the Portuguese Foundation for Science and Technology (FCT) under the contracts: PTDC/FISMAC/5689/2020, EXPL/FIS-MAC/0406/2021, UIDB/00618/2020 and UIDP/00618/2020.


%

\end{document}